\documentclass[%
,twocolumn%
,secnumarabic%
,amssymb,aps,pra,nobibnotes]{revtex4-1}
\usepackage{srcltx}
\usepackage{psfrag}
\usepackage{subfigure}
\usepackage{epsfig}
\usepackage{amsmath}
\usepackage[usenames]{color}
\expandafter\ifx\csname package@font\endcsname\relax\else
 \expandafter\expandafter
 \expandafter\usepackage
 \expandafter\expandafter
 \expandafter{\csname package@font\endcsname}%
\fi
\DeclareRobustCommand\substyle{\name@idx{document substyle}}%
\DeclareRobustCommand\classoption{\name@idx{document class option}}%
\DeclareRobustCommand\classname{\name@idx{document class}}%
\def\name@idx#1#2{%
 {\ttfamily#2}%
 \index{#2\space#1=\string\ttt{#2}\space#1}\index{#1>#2=\string\ttt{#2}}%
}%

\begin{document}
\title{Perfect Dispersive Medium}%
\author{Shulabh Gupta and Christophe Caloz}%
\affiliation{\'{E}cole Polytechnique de Montr\'{e}al, Montr\'{e}al, Qu\'{e}bec, H3T 1J4, Canada (e-mail: shulabh.gupta@polymtl.ca).}

\begin{abstract}
Dispersion lies at the heart of real-time signal processing systems across the entire electromagnetic spectrum from radio to optics. However, the performance and applicability of such systems have been severely plagued by distortions due to the frequency dependent nature of the amplitude response of the dispersive media used for processing. This frequency dependence is a fundamental consequence of the causality constraint, incarnated by Kramers-Kronig relations or, equivalently, by the Bode relations. In order to resolve this issue, we introduce here the concept of a \emph{perfect dispersive medium}, which is a loss-gain medium characterized by a perfectly flat magnitude response along with an arbitrary phase response. This unprecedented property results from equalized electric and magnetic dipole dispersion responses, whence the amplitude and phase of the transmission functions of the isolated loss and gain contributions become the inverse and remain the same, respectively, under reversal of the sign of the imaginary part of the equalized magneto-dielectric polarizability. Such a perfect dispersive medium may be realized in the form of a metamaterial, and the paper demonstrates a corresponding stacked loss-gain metasurface structure for illustration. From a practical standpoint, perfect dispersive media represent a paradigm shift that may propel real-time signal processing technology to a new dimension, with a myriad of novel ultrafast communication, sensing, imaging and instrumentation applications.
\end{abstract}
\maketitle

\section{Introduction}

\begin{figure}[htbp]
\begin{center}
\subfigure[]{
\psfrag{d}[c][c][1]{$\psi_\text{in}(\omega)$}
\psfrag{e}[c][c][1]{$\psi_\text{out}(\omega)$}
\psfrag{f}[c][c][1]{$\psi_\text{ref}(\omega)$}
\psfrag{t}[c][c][1]{$t$}
\psfrag{i}[c][c][1]{$\Delta\tau$}
\includegraphics[width=0.75\columnwidth]{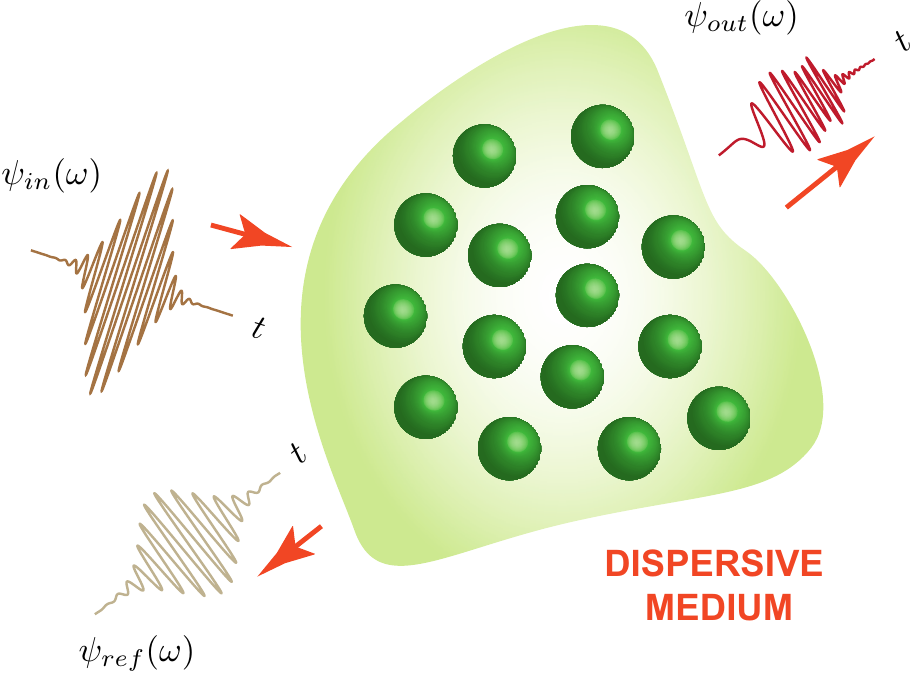}}
\subfigure[]{
\psfrag{a}[c][c][1]{$\omega$}
\psfrag{x}[c][c][0.9]{\eqref{Eq:KK} and \eqref{Eq:Bode_Phase_Amplitude}}
\psfrag{b}[l][c][1]{$\tau(\omega)$}
\psfrag{c}[c][c][1]{$T(\omega)$}
\psfrag{i}[c][c][1]{$\Delta\tau$}
\psfrag{g}[c][c][1]{$\tau_1(\omega)$}
\psfrag{h}[c][c][1]{$\tau_2(\omega)$}
\psfrag{j}[c][c][1]{$\tau_3(\omega)$}
\psfrag{n}[c][c][1]{$T_1(\omega)$}
\psfrag{p}[c][c][1]{$T_2(\omega)$}
\psfrag{m}[c][c][1]{$T_3(\omega)$}
\psfrag{q}[c][c][0.75]{$T_1(\omega)\ne T_2(\omega)\ne T_3(\omega)\ne 1$}
\includegraphics[width=\columnwidth]{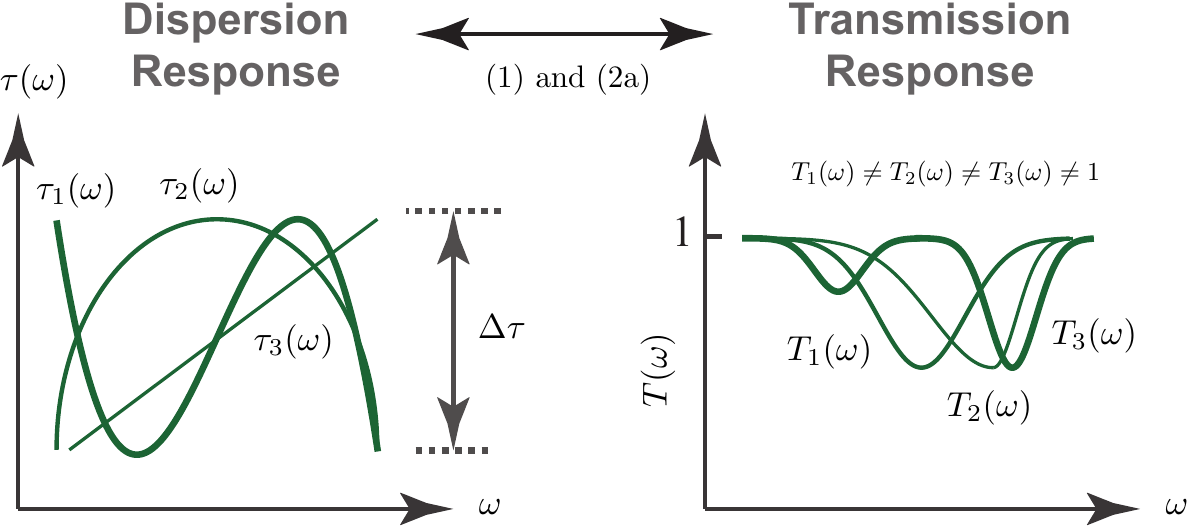}}
\subfigure[]{
\psfrag{D}[c][c][1]{$\Delta z >> \lambda_0$}
\psfrag{E}[c][c][1]{$\delta \ll \lambda_0$}
\psfrag{F}[c][c][1]{$\{\hat{\alpha}_\text{e}, \hat{\alpha}_\text{m}\}$}
\psfrag{G}[c][c][1]{$R(\omega)$}
\psfrag{H}[c][c][1]{$T(\omega)$}
\psfrag{X}[c][c][1]{$E_x$}
\psfrag{Y}[c][c][1]{$H_y$}
\psfrag{Z}[c][c][1]{$k$}
\psfrag{a}[c][c][1]{$\omega$}
\psfrag{c}[c][c][1]{$T(\omega)$}
\psfrag{r}[c][c][1]{$T_m(\omega)= 1,\forall m$}
\includegraphics[width=\columnwidth]{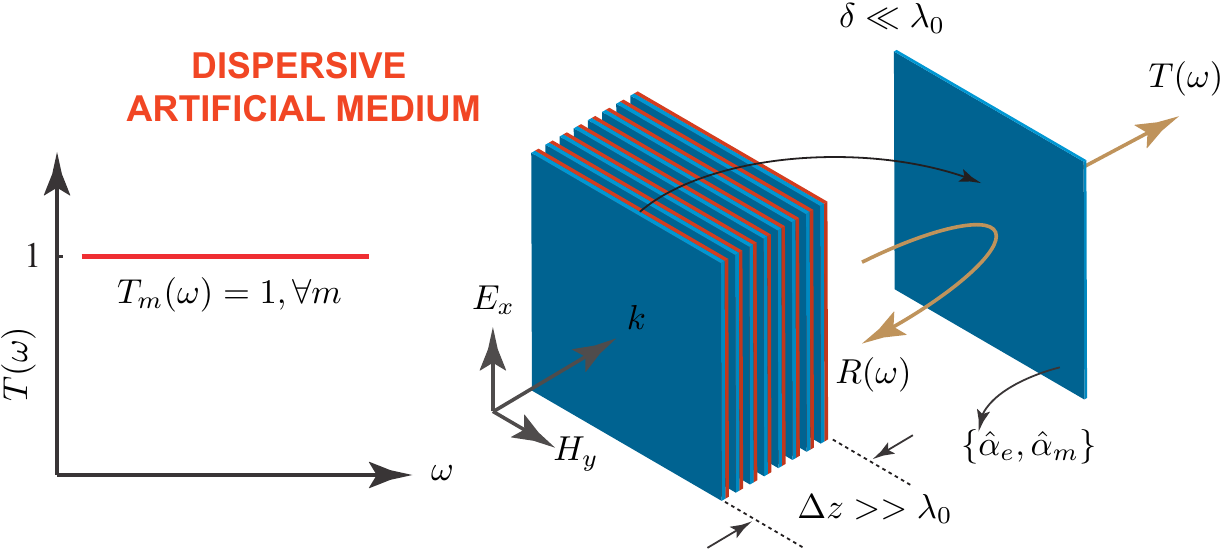}}
\caption{Issues in natural dispersive media and proposed perfect dispersive medium solution. a)~General dispersive medium scattering light. b)~Conventional dispersion, with frequency-dependent group delay and magnitude responses following Kramers-Kronig and Bode relations. c)~Perfect dispersive medium, implemented here in the form of stacked loss-gain metasurfaces.}
\label{Fig:Definition}
\end{center}
\end{figure}

All media, except vacuum, are dispersive. While dispersion often represents an undesired effect, it is the phenomenon underpinning ultrafast signal processing systems across the entire electromagnetic spectrum from radio to optical frequencies~\cite{Jour:2013_MwMag_Caloz,Dragoman_THZ,THz_processing,Goodman_Fourier_Optics}. Dispersion provides a natural mechanism to manipulate temporal frequencies ($\omega$) in the pristine analog form of electromagnetic waves and a diversity of related real-time signal processing devices and systems have been developed in recent decades. A dispersive medium modulates the instantaneous frequency contents of broadband pulses, as illustrated in Fig.~\ref{Fig:Definition}(a), and, when properly controlled, this effect enables devices such as pulse shapers, real-time Fourier transformers, and dispersion compensators, to name a few \cite{Jannson_OL_04_1983,Agrawal_NLFO_1980}.

Unfortunately, in a naturally occurring dispersive medium, that is here assumed to be both passive and linear, causality imposes fundamental restrictions on the achievable dispersion responses. These restrictions are expressed by the Kramers-Kronig relations, that relate the real and imaginary parts of the medium parameters as

\begin{subequations}
\begin{equation}
 \chi''(\omega)=\frac{2\omega}{\pi}\mathcal{P}\int_{0}^{+\infty}\frac{\chi'(\omega') }{\omega^{'2} - \omega^2} d\omega',\\
\end{equation}
\begin{equation}
\chi'(\omega) = -\frac{2}{\pi}\mathcal{P} \int_{0}^{+\infty}\frac{\omega' \chi{''}(\omega') }{\omega^{'2} - \omega^2} d\omega',
\end{equation}\label{Eq:KK}
\end{subequations}

\noindent where $\chi(\omega)$ could, for instance, represent the electric susceptibility or the magnetic susceptibility of the medium~\cite{KK_Active}. These relations may alternatively be written in terms of the magnitude and phase of the transmission transfer function, $T(\omega)$, as~\cite{KK_Bode_Zeros}

\begin{subequations}\label{Eq:Bode_rel_pair}
\begin{equation}
\angle T(\omega) = -\frac{2\omega}{\pi}\mathcal{P}\int\displaylimits_0^{+\infty}  \frac{\ln |T(\omega')|}{\omega'^2 - \omega^2}d\omega' + \frac{\omega z}{c},\label{Eq:Bode_Phase_Amplitude}
\end{equation}
\begin{equation}
\ln |T(\omega)|=\frac{2\omega}{\pi}\mathcal{P}\int\displaylimits_0^{+\infty}\frac{\angle T(\omega') - (\omega' z/c)}{\omega'^2 - \omega^2}d\omega'.
\end{equation}
\end{subequations}

\noindent These relations are called the Bode relations and are illustrated in Fig.~\ref{Fig:Definition}(b) for some conventional dispersive examples. They are obtained by taking the logarithm of the propagating waveform, $T(\omega,z)=e^{jkz}=e^{jn_0(\omega)z/c}$, where $n_0(\omega)=n(\omega)+i\kappa(\omega)$ is the complex refractive index of the medium, and substituting the result in the refractive index counterpart of~\eqref{Eq:KK}~\cite{Jackson_book_CED}. In addition, a linear phase term, $\omega z/c$, has been included to account for phase accumulation through eventual non-dispersive (air) sections of the medium between its ``atoms''.

However, the relations~\eqref{Eq:Bode_rel_pair} are not completely general. If $T(\omega)$ includes transmission zeros, the formulation~\eqref{Eq:Bode_rel_pair} is not possible due to consequent singularities associated with the logarithm function, and in this case the magnitude-to-phase relationship~\eqref{Eq:Bode_Phase_Amplitude}, which will be later needed in the paper, takes the more general form~[see Supplementary Material, Sec.~I]

\begin{equation}
\begin{split}
\angle T(\omega) = &-\frac{2\omega}{\pi}\mathcal{P} \int\displaylimits_0^{+\infty}\frac{ \ln |T(\omega')|}{\omega'^2 - \omega^2} d\omega' + \frac{\omega z}{c}\\ &+ \sum_{n=1}^N\angle\left(\frac{\omega - \omega_n}{\omega_n^\ast - \omega}\right) ,\label{Eq:Bode_MS}
\end{split}
\end{equation}

\noindent where $N$ is the number of transmission zeros, $\omega_n$, in the upper half of the complex $\omega$ plane, causally corresponding to the lower-plane transmission poles $\omega_n^*$.

According to \eqref{Eq:Bode_Phase_Amplitude}, a flat transmission magnitude response, $|T|=\text{const.}$, is incompatible with dispersion for such a response entails a purely linear phase~\cite{KK_Active}, $\angle T(\omega)\propto\omega$, and hence a constant (non-dispersive) group delay response. Conversely, dispersion, characterized by $\angle T(\omega)$ being a nonlinear function of frequency or, equivalently, the group delay being frequency dependent, implies amplitude frequency variation, $|T|=|T(\omega)|$, which unfortunately induces distortions that limit the performance and functionality of dispersion-based signal processing systems~\cite{Jour:2013_MwMag_Caloz,Dragoman_THZ,THz_processing}.

To overcome this fundamental issue, we introduce here the concept of a perfect dispersive medium, which is a medium exhibiting an arbitrary phase versus frequency response along with a perfectly flat magnitude over the entire spectrum, as illustrated in Fig.~\ref{Fig:Definition}(c). Such a medium may be implemented in the form of a loss-gain metamaterial structure with equalized electric and magnetic dipole Lorentz responses in both the loss and gain contributions. In this medium, the Kramers-Kronig relations, that represent a completely general statement of causality, will be shown to be still satisfied, but in an unusual fashion, whereby $\chi''(\omega)$ reduces to a set of Dirac delta functions that is unrelated to loss given its zero bandwidth, while the Bode relation~\eqref{Eq:Bode_MS} will be shown to reduce to its last two terms.

\section{Formation of Perfect Dispersive Metamaterial Structure}\label{eq:form_perf_disp_mtm}

We will show here that, although not existing in nature, a perfect dispersive medium may be realized under the form of a loss-gain artificial material or metamaterial~\cite{Caloz_Wiley_2006,MTM_Eleftheriades,Jour:Smith_MTM}. A metamaterial can be generally characterized in terms of volumetric electric and magnetic dipolar moments, $\mathbf{p}$ and $\mathbf{m}$, respectively, modeling the constituent scattering particles. For the sake of the argument, but without loss of generality, we shall consider here the metamaterial implementation depicted in Fig.~\ref{Fig:Definition}(c), which consists of stacked sub-wavelengthly thick films incorporating scattering particles, or metasurfaces \cite{Metasurface_Review,Capasso_metasurfaces}.

A metasurface may be modeled by surface polarizabilities, that relate the local electromagnetic fields to the induced dipole moments. Specifically, the relations read $\mathbf{p} = \hat{\alpha}_\text{e}\mathbf{E_\text{loc}}$ and $\mathbf{m} = \hat{\alpha}_\text{m}\mathbf{H_\text{loc}}$, where $\mathbf{E_\text{loc}}$ and $\mathbf{H_\text{loc}}$ are the electric and magnetic local fields, respectively, and $\hat{\alpha}_\text{e}$ and $\hat{\alpha}_\text{m}$ are the effective electric and magnetic surface polarizabilities~\cite{Grbic_Metasurfaces}, respectively, including the effect of mutual coupling. It is assumed that $\mathbf{p} $ and $\mathbf{m}$ are oriented perpendicular to each other and in the plane of the metasurface, a configuration that is commonly referred to as the Huygens' source and that naturally occurs in the case of plane wave normal incidence on a metasurface. In such a situation, the reflection and transmission functions of the metasurface are related to the surface polarizabilities as~\cite{Metasurface_Synthesis_Caloz,Grbic_Metasurfaces}

\begin{subequations}
\begin{equation}
R(\omega)
=\frac{2i\omega\left[\eta_0^2\hat{\alpha}_\text{e}(\omega) - \hat{\alpha}_\text{m}(\omega)\right]}{\left[ 2\eta_0-i\omega\hat{\alpha}_\text{m}(\omega)\right]\left[2 - i\omega\eta_0\hat{\alpha}_\text{e}(\omega)\right]},\label{Eq:Romega}
\end{equation}
\begin{equation}
T(\omega)
=\frac{\left[4+\omega^2\hat{\alpha}_\text{e}(\omega)\hat{\alpha}_\text{m}(\omega)\right]\eta_0}{\left[\omega\hat{\alpha}_\text{m}(\omega) +2i\eta_0\right]\left[\eta_0\omega\hat{\alpha}_\text{e}(\omega) +2i\right]},
\end{equation}\label{Eq:Tomega}
\end{subequations}

\noindent where $\eta_0$ is the impedance of free space.

Equation~\eqref{Eq:Romega} reveals that reflection is perfectly vanishes under the condition

\begin{equation}\label{eq:matching_cond}
\hat{\alpha}_\text{m}(\omega)
= \eta_0^2\hat{\alpha}_\text{e}(\omega).
\end{equation}

\noindent In practice, this condition is easily satisfied over the entire frequency spectrum ($\forall~\omega$), due to naturally compatible causal electric and magnetic relations, for instance by equalizing the resonance and antiresonance frequencies of the electric and magnetic responses in a Lorentz medium. Note that providing complete reflection suppression, or perfect matching, is certainly a useful capability for a metasurface, and that this capability is generally not available in natural materials.

Under the reflection-less condition~\eqref{eq:matching_cond}, the transmission function~\eqref{Eq:Tomega} reduces to

\begin{equation}\label{eq:Tomega}
T_\text{m}(\omega) = \frac{1 + i\eta_0\omega\hat{\alpha}_\text{e}(\omega)/2}{1 - i\eta_0\omega\hat{\alpha}_\text{e}(\omega)/2},
\end{equation}

\noindent where the subscript ``m'' emphasizes the fact that this expression corresponds to the particular case of a matched metasurface, and it will next be shown that this transfer function describes a perfect dispersive medium.

\begin{figure*}[htbp]
\begin{center}
\psfrag{b}[c][c][1]{Frequency (GHz)}
\psfrag{a}[c][c][1]{$|T_{xx}|(\omega)$ (dB)}
\psfrag{c}[c][c][1]{$\tau(\omega)$ (ns)}
\psfrag{d}[c][c][1]{$\eta_0\hat{\alpha}_\text{e}(\omega)$, $\hat{\alpha}_\text{m}(\omega)/\eta_0$}
\psfrag{e}[c][c][0.8]{$\mathbf{p}$}
\psfrag{f}[c][c][0.8]{$\mathbf{m}$}
\psfrag{g}[c][c][1]{$|R_{xx}(\omega)|=0$}
\psfrag{h}[c][c][1]{${\gamma> 0}$}
\psfrag{j}[c][c][1]{${\gamma< 0}$}
\psfrag{w}[c][c][0.8]{$E_{x}$}
\psfrag{k}[c][c][1]{$\tau(\gamma) = \tau(-\gamma)$}
\psfrag{m}[c][c][0.8]{$\Delta\tau$}
\psfrag{n}[c][c][0.8]{$2\Delta\tau$}
\psfrag{p}[c][c][1]{$|T_{\text{m}{xx}}(\omega) = 1$}
\psfrag{q}[c][c][1]{$|R_{\text{m}{xx}}|(\omega) = 0$}
\psfrag{x}[c][c][1]{$x$}
\psfrag{y}[c][c][1]{$y$}
\psfrag{z}[c][c][1]{$z$}
\psfrag{A}[l][c][0.7]{Re$\{\cdot\}$}
\psfrag{B}[l][c][0.7]{Im$\{\cdot\}$}
\includegraphics[width=2\columnwidth]{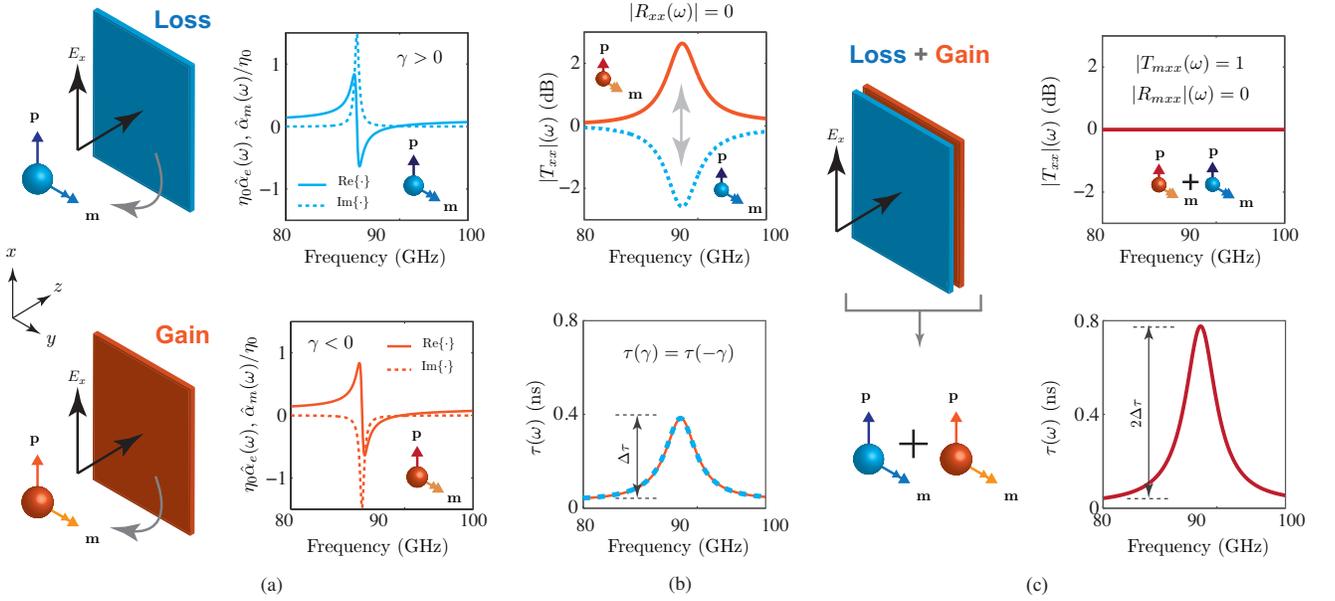}
\caption{Numerical demonstration of a perfect dispersive metasurface formed a stacked loss-gain metasurface structure. a) Lorentzian dispersion profiles of the dispersion equalized (or matched) electric and magnetic polarizabilities for the loss or gain cases. b) Corresponding transmission and group delay delay responses. c) Transmission and a delay responses of the composite loss-gain metasurface. Reflection is zero over the entire frequency spectrum in all cases.}
\label{Fig:Lorentzmedium}
\end{center}
\end{figure*}

Let us first examine the simple case of a hypothetical lossless medium. In such a medium, $\hat{\alpha}_\text{e}(\omega)$ is a purely real function, and substitution of a real variable $\hat{\alpha}_\text{e}$ into~\eqref{eq:Tomega} leads to the result $|T(\omega)|=1$, consistently with energy conservation. In this case, the transfer function takes the form

\begin{equation}
T_\text{m}(\omega)=\frac{G(\omega)}{G^\ast(\omega)},\label{Eq:Ap}
\end{equation}

\noindent with $G(\omega) = 1 + i\eta_0\omega\hat{\alpha}_\text{e}/2$, which is the transfer function of a system commonly referred to as an \emph{all-pass} system in electronics engineering~\cite{MYJ_AHP_1980,Oppenheim_Schafer_DTSP}. However, such a system is only an idealization of a low-loss dispersive system; it is strictly unphysical since any physical system includes loss.

Let us therefore consider now a physical, and hence lossy, medium. In such a medium, the polarizability is complex, i.e. the term $\hat{\alpha}_\text{e}(\omega)/2$ in \eqref{eq:Tomega} reads $\hat{\alpha}_\text{e}(\omega)/2 = \hat{\alpha}_r(\omega) + i\hat{\alpha}_i(\omega)$, where $\hat{\alpha}_i(\omega) > 0$ and $\hat{\alpha}_i(\omega) < 0$ correspond to loss and gain, respectively, assuming the phasor convention $e^{-i\omega t}$. In this case, the matched transmission function~\eqref{eq:Tomega} reads

\begin{equation}
T_\text{m}(\omega) = \frac{[1 - \eta_0\omega\hat{\alpha}_i(\omega)] + i \eta_0\omega\hat{\alpha}_r(\omega)}{[1 + \eta_0\omega\hat{\alpha}_i(\omega)] - i \eta_0\omega\hat{\alpha}_r(\omega)},\label{Eq:TF_Physical}
\end{equation}

\noindent and exhibits the phase response

\begin{subequations}\label{Eq:Tmagphase}
\begin{equation}
\angle T_\text{m}(\omega) = \tan^{-1}\left(\frac{\eta_0\omega\hat{\alpha}_r }{1 - \eta_0\omega\hat{\alpha}_i}\right) + \tan^{-1}\left(\frac{\eta_0\omega\hat{\alpha}_r }{1 + \eta_0\omega\hat{\alpha}_i}\right)\label{Eq:Tphase}
\end{equation}
and the magnitude response
\begin{equation}
|T_\text{m}(\omega)|=\sqrt{\frac{(1 - \eta_0\omega\hat{\alpha}_i)^2 + (\eta_0\omega\hat{\alpha}_r)^2 }{(1 + \eta_0\omega\hat{\alpha}_i)^2 + (\eta_0\omega\hat{\alpha}_r)^2}}.\label{Eq:Tmagnitude}
\end{equation}
\end{subequations}

It follows from~\eqref{Eq:Tmagphase} that

\begin{subequations}\label{Eq:Keyphasemag}
\begin{equation}\label{Eq:Keyphase}
\angle T_\text{m}(\omega, \hat{\alpha}_i(\omega)) = \angle T(\omega, -\hat{\alpha}_i(\omega)) = \phi(\omega)
\end{equation}
and
\begin{equation}\label{Eq:Key}
|T_\text{m}(\omega,\hat{\alpha}_i(\omega))| = \frac{1}{|T(\omega,-\hat{\alpha}_i(\omega))|} < 1,
\end{equation}
\end{subequations}

\noindent which reveals that matched loss and gain metasurfaces are perfectly equivalent and complementary to each other in terms of phase and magnitude, respectively, over the entire frequency spectrum. Thus, a composite system obtained by pairing a loss $(\hat{\alpha}_i > 0)$ metasurface and a gain $(\hat{\alpha}_i > 0)$ metasurface satisfying both the conditions~\eqref{eq:matching_cond} forms a \emph{perfect dispersive surface}~(PDS), while a 3D medium obtained by cascading composite loss-gain metasurfaces (Fig.~\ref{Fig:Definition}(c)) forms a \emph{perfect dispersive material}~(PDM). A perfect dispersive surface is characterized by the transmission function

\begin{equation}
T_\text{PDS}(\omega)  =  \overbrace{T_\text{m}(\omega,\hat{\alpha}_i(\omega))}^{\text{Loss}}\times \overbrace{T_\text{m}(\omega,-\hat{\alpha}_i(\omega))}^{\text{Gain}},\label{Eq:PDS}
\end{equation}

where, according to~\eqref{Eq:Keyphasemag},

\begin{subequations}\label{Eq:TF_MTM}
\begin{equation}\label{Eq:TF_MTM_mag}
|T_\text{PDS}(\omega)|=1
\end{equation}
and
\begin{equation}
\angle T_\text{PDS}(\omega) = \exp\{i2\phi(\omega)\}.
\end{equation}
\end{subequations}

Substituting~\eqref{Eq:TF_MTM_mag} into the general Bode relation~\eqref{Eq:Bode_MS} and considering the fact that the  metasurface thickness is zero, demanding $\omega z/c=0$, yields

\begin{equation}
\angle T_\text{PDS}(\omega)
=\angle\left(\frac{\omega - \omega_n}{\omega_n^\ast - \omega}\right),
\end{equation}

\noindent where the transmission magnitude term has vanished and where it appears that dispersion is exclusively produced by transmission zeros. This results represents the proof that perfect dispersion is not at odds with causality and naturally occurs in a (matched) composite loss-gain medium with conjugately related transmission zeros and poles. Similarly, for a perfect dispersive material, formed by periodically cascading $N$ pairs of loss-gain metasurfaces with a period $p$, the above Bode relations transforms to

\begin{equation}
\angle T_\text{PDM}(\omega) =  \sum_{n=1}^N\angle\left(\frac{\omega - \omega_n}{\omega_n^\ast - \omega}\right) + \frac{\omega pN}{c}.
\end{equation}

\section{Particular Case of a Lorentz Perfect Dispersive Metamaterial}\label{sec:part_Lorentz_PDM}

To demonstrate the feasibility of a perfect dispersive medium, let us consider a metasurface made of scattering particles having a Lorentz dispersive response \cite{Jackson_book_CED}. Such a response reads

\begin{equation}
\hat{\alpha}(\omega,\gamma) = \frac{A\omega_p^2}{(\omega_0^2-\omega^2) - i\gamma\omega},\label{Eq:Lorentz_dipolar_moment}
\end{equation}

\noindent where $\omega_p$ and $\omega_0$ are the plasma and resonant frequencies, respectively, and $\gamma > 0$ and $\gamma < 0$ represent loss or gain, respectively, and is plotted in Fig.~\ref{Fig:Lorentzmedium}(a) for isolated loss and gain media with parameters equalized according to~\eqref{eq:matching_cond}.

Substituting \eqref{Eq:Lorentz_dipolar_moment} with \eqref{eq:matching_cond} into~\eqref{eq:Tomega} yields the corresponding  transfer function

\begin{equation}
T_\text{m}^\text{Lorentz}(\omega,\gamma)  = \frac{(2\gamma - A\eta_0\omega_p^2) \omega + 2i(\omega_0^2-\omega^2)}{(2\gamma + A\eta_0\omega_p^2) \omega + 2i(\omega_0^2-\omega^2)}. \label{Eq:TFLp}
\end{equation}

\noindent The causality of this function depends on the sign of $\gamma$. In a lossy metasurface ($\gamma > 0$), the transmission poles are always located in the lower half of the complex plane so that the transfer function is unconditionally causal. In contrast, in a gain metasurface ($\gamma < 0$) causality requires the condition $|\gamma|<\eta_0\omega_p^2/2$, placing all the transmission poles of~\eqref{Eq:TFLp} in the lower half of the complex plane, to be specifically satisfied.

Figure~\ref{Fig:Lorentzmedium}(b) plots the Lorentz metasurface transfer function using~\eqref{Eq:TFLp} for $\gamma>0$ and $\gamma<0$ as well as the corresponding group delays. As expected from the general properties~\eqref{Eq:Keyphasemag} for matched metasurfaces, Lorentz loss and gain metasurfaces exhibit perfectly symmetrical magnitude responses and identical group delay response. Both metasurfaces, in isolation, represent poor dispersive media since their dispersion is achieved at the cost of magnitude distortion according to the Bode relation~\eqref{Eq:Bode_MS}. Note that, because $|T_\text{m}(\omega)|\ne 1$ in either case, both metasurfaces have their dispersion responses depending on the transmission magnitude, since we have then $\ln|T_\text{m}(\omega)|\neq 0$ in the integral term of~\eqref{Eq:Bode_MS}. In addition, their dispersion responses naturally also depend on the zeros lying in the upper half of the complex plane, corresponding to the third term of~\eqref{Eq:Bode_MS}.

The transmission function of the composite Lorentz loss-gain metasurface formed by stacking together the loss and gain metasurfaces in Fig.~\ref{Fig:Lorentzmedium}(b) is obtained, according to~\eqref{Eq:PDS}, as

\begin{equation}
T_\text{PDS}^\text{Lorentz}(\omega)  = T_\text{m}^\text{Lorentz}(\omega, \gamma)\times T_\text{m}^\text{Lorentz}(\omega, -\gamma).
\label{Eq:TPDSL}
\end{equation}

\noindent This response is plotted in Fig.~\ref{Fig:Lorentzmedium}(c), using~\eqref{Eq:TFLp} into~\eqref{Eq:TPDSL}. As predicted by \eqref{Eq:TF_MTM}, the transmission and reflection magnitudes of this metasurface are perfectly flat and zero, respectively, over the entire frequency spectrum, while the group delay response is twice that of the isolated loss or gain metasurfaces. The composite structure is thus a PDS.

The composite loss-gain metasurface may be described in terms of its effective surface polarizability, $\hat{\alpha}_\text{eff}(\omega)$, given by [Supplementary Material, Sec. II],

\begin{subequations}
\begin{equation}
\begin{split}
\hat{\alpha}_\text{eff}^\text{Lorentz}(\omega) &= \left[\frac{c_1}{\omega_1^2 - \omega^2}-i\pi\{\delta(\omega+\omega_1) - \delta(\omega-\omega_1)\}\right] +  \\
&  \left[\frac{c_2}{\omega_2^2 - \omega^2}-i\pi\{\delta(\omega+\omega_2) - \delta(\omega-\omega_2)\}\right],	\label{Eq:PolEffLossGain}
\end{split}
\end{equation}
\begin{equation}
\text{with}\quad\omega_{1,2}^2 = \frac{\kappa \pm 8\sqrt{\kappa^2 -\omega_0^4}}{8},
\end{equation}
\end{subequations}

\noindent where $\kappa = (A^2\eta_0^2\omega_p^2 + 8\omega_0^2+ 4\gamma^2)/8$, $c_1= 8A\omega_p^2(\omega_1^2-\omega_0^2)/(\omega_1^2-\omega_2^2)$ and  $c_2= 8A\omega_p^2(\omega_0^2-\omega_2^2)/(\omega_1^2-\omega_2^2)$. It can be easily verified that $\hat{\alpha}_\text{eff}(\omega)$ satisfies the Kramers-Kronig relations~\eqref{Eq:KK} and thus represents a causal polarization response. Furthermore, it may be verified that substituting~\eqref{Eq:PolEffLossGain} into~\eqref{eq:Tomega} leads to $|T_\text{m}(\omega)|=1$. Consequently, since $\ln|T_\text{m}(\omega)|=0$, the corresponding dispersion response is independent of the transmission magnitude (integral term in~\eqref{Eq:Bode_MS}) and depends only on the transmission zeros (third term in~\eqref{Eq:Bode_MS}).

\begin{figure}[htbp]
\begin{center}
\psfrag{a}[c][c][0.8]{Frequency (GHz)}
\psfrag{b}[c][c][0.8]{normalized group delay, $\tau(\omega) - \text{min}\{\tau(\omega)\}$ (ns)}
\psfrag{c}[c][c][0.8]{Transmission $|T_{xx}(\omega)|$ (dB)}
\psfrag{d}[l][c][0.7]{spec.}
\psfrag{e}[l][c][0.7]{design}
\psfrag{f}[l][c][0.7]{loss-gain}
\psfrag{h}[l][c][0.7]{loss-loss}
\psfrag{g}[c][c][0.8]{pair index $n$}
\psfrag{i}[c][c][0.7]{$f_0$}
\psfrag{j}[c][c][0.7]{$f_p$}
\psfrag{k}[c][c][0.8]{frequency $\omega_p/2\pi$, $\omega_0/2\pi$ (GHz)}
\psfrag{A}[c][c][0.8]{$n=1$}
\psfrag{B}[c][c][0.8]{$n=2$}
\psfrag{C}[c][c][0.8]{$n=3$}
\psfrag{D}[c][c][0.8]{$n=N$}
\psfrag{E}[c][c][0.8]{$\{\omega_{0,n}, \omega_{p,n}\}$}
\includegraphics[width=1\columnwidth]{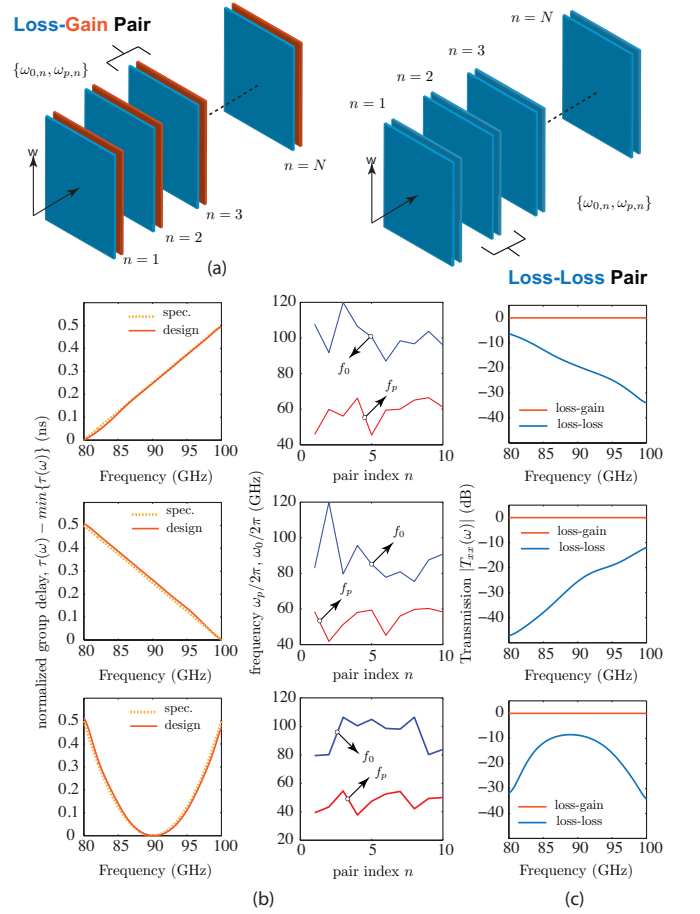}
\caption{Group delay engineering with a perfect dispersive metamaterial composed of stacked perfect dispersive Lorentz composite loss-gain metasurfaces. a)~Metamaterial (left), compared with its purely lossy counterpart (right). (b)~Group delay responses realized with \eqref{Eq:GDE} (left) with the composite layers resonance-frequency parameters plotted in the right. c)~Magnitude transmission responses, compared to those of a purely loss structure. In all cases, $A = 3\times10^{-15}$, $\gamma=12\times10^{-9}$ and $N=10$.}
\label{Fig:MetasurfaceCascade}
\end{center}
\end{figure}

From this point, a perfect dispersive metamaterial with \emph{arbitrary delay response}~\cite{Gupta_TMTT_09_2010} can be designed by cascading perfect dispersive composite loss-gain metasurfaces with different resonance and plasma frequency parameters, as illustrated in Fig.~\ref{Fig:MetasurfaceCascade}(a). The transfer function of such a metamaterial may be expressed as

\begin{subequations}
\begin{equation}
T(\omega) = \prod_{n=1}^{N}\overbrace{\left(\frac{1 +i\eta_0\omega\hat{\alpha}_n/2}{1 - i\eta_0\omega\hat{\alpha}_n/2}\right)}^{\text{Loss}}\overbrace{\left(\frac{1 + i\eta_0\omega\hat{\alpha}^{\ast}_n/2}{1 - i\eta_0\omega\hat{\alpha}^{\ast}_n/2}\right)}^{\text{Gain}}\label{Eq:GDEa}
\end{equation}
\begin{equation}
\text{with}\quad\hat{\alpha}_n=\hat{\alpha}_n(\omega) = \frac{A\omega_{p,n}^2}{(\omega_{0, n}^2-\omega^2) - i\gamma\omega}.
\end{equation}\label{Eq:GDE}
\end{subequations}

\noindent Figure~\ref{Fig:MetasurfaceCascade}(b) plots various -- linear up-chirp, linear down-chirp and quadratic -- group delay responses achieved in this manner, using~\eqref{Eq:GDE}. As expected, the magnitude transmission response is flat and
$0-$dB in all cases. This is in contrast with the conventional composite loss-loss metasurfaces, where the magnitude transmission strongly depends on dispersion, as shown in Fig.~\ref{Fig:MetasurfaceCascade}(c).

\begin{figure*}[htbp]
\begin{center}
\psfrag{a}[c][c][1]{frequency (GHz)}
\psfrag{b}[c][c][1]{$T_{xx}$, $R_{xx}$ (dB)}
\psfrag{c}[c][c][1]{group delay (ps)}
\psfrag{d}[c][c][1]{$\Delta\tau_{1}$}
\psfrag{e}[c][c][1]{$\Delta\tau_{2}$}
\psfrag{f}[c][c][1]{$\Delta\tau_{3}$}
\psfrag{g}[c][c][1]{$\delta$}
\psfrag{h}[c][c][0.9]{$-5.5$ dB}
\psfrag{i}[c][c][0.9]{$+5.5$ dB}
\psfrag{x}[c][c][0.9]{$x$}
\psfrag{y}[c][c][0.9]{$y$}
\psfrag{z}[c][c][0.9]{$z$}
\psfrag{j}[c][c][0.9]{$E_{x}$}
\psfrag{k}[c][c][0.9]{$\boxed{\Delta\tau_1 = \Delta\tau_2\quad |T_{xx}^G| = -|T_{xx}^L|}$}
\psfrag{m}[c][c][0.9]{$\boxed{\Delta\tau_3 = 2\Delta\tau_{1,2}\quad |T_{xx}| \approx 0\text{dB}}$}
\psfrag{l}[c][c][0.9]{$a$}
\psfrag{n}[c][c][0.9]{$\Delta x$}
\psfrag{o}[c][c][0.9]{$\Delta y$}
\psfrag{p}[c][c][0.9]{$s_0$}
\psfrag{q}[c][c][0.9]{$h$}
\psfrag{r}[c][c][0.9]{$\varepsilon_r$}
\psfrag{s}[c][c][0.9]{$\varepsilon_r^\ast$}
\psfrag{t}[c][c][0.9]{$\varepsilon_r = 1$}
\includegraphics[width=2\columnwidth]{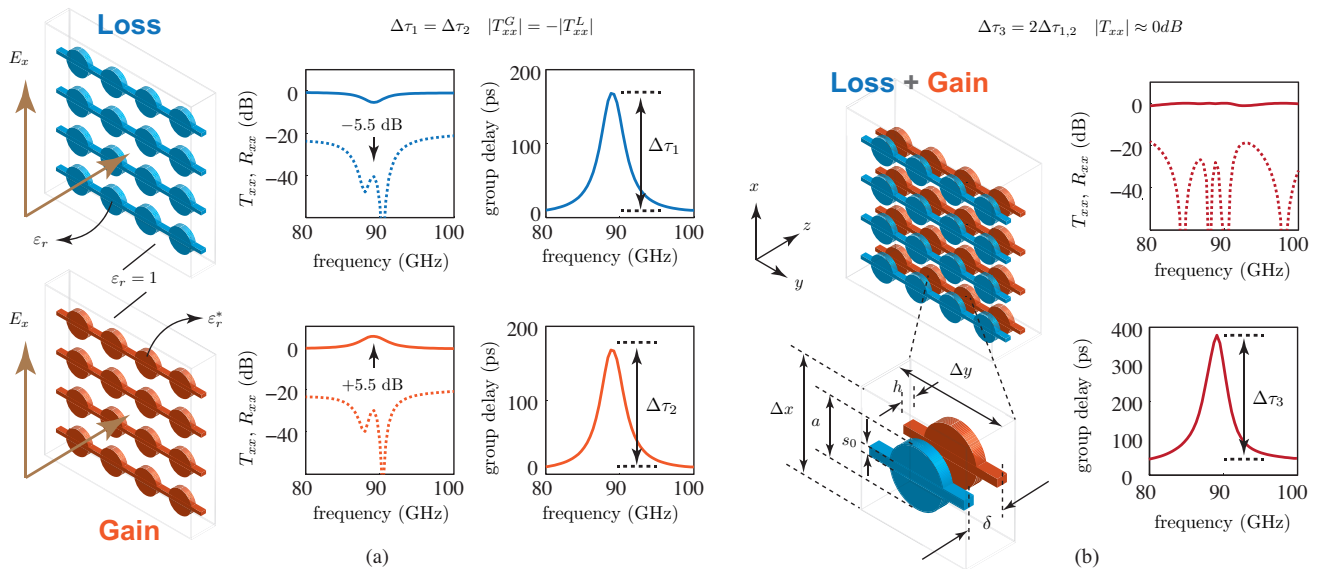}
\caption{Full-wave demonstration of a Lorentz perfect dispersive metasurface formed by a 2D array of circular cylindrical sub-wavelength circular-cylindrical dielectric resonators. a)~Isolated loss and gain metasurfaces (left) and their transmission/reflection and delay responses (right). b)~Composite loss-gain metasurface (left) and its transmission/reflection and delay responses (right). All the results are computed using the Rigorous Coupled Wave Analysis (RCWA) technique. The parameters are $\Delta x = \Delta y = 2.65~$mm, $a=h=0.635$~mm, $\delta = 3.175$~mm, $s_0 = 0.508$~mm, $\varepsilon_r = \varepsilon^{\prime} + i\varepsilon^{\prime\prime} = 10.2 + i1.02$, number of space harmonics used in RCWA to ensure convergence $= 21$.}
\label{Fig:Metasurface}
\end{center}
\end{figure*}

\begin{figure}[htbp]
\begin{center}
\psfrag{a}[c][c][1]{frequency (GHz)}
\psfrag{b}[c][c][1]{$T_{xx}$, $R_{xx}$ (dB)}
\psfrag{c}[c][c][1]{$\tau(\omega)$ (ns)}
\psfrag{d}[c][c][0.8]{$\Delta\tau_{1}$}
\psfrag{e}[c][c][0.8]{$\Delta\tau_{2}$}
\psfrag{f}[c][c][0.8]{$\Delta\tau$}
\psfrag{g}[c][c][0.8]{$\delta$}
\psfrag{x}[c][c][1]{$x$}
\psfrag{y}[c][c][1]{$y$}
\psfrag{z}[c][c][1]{$z$}
\psfrag{j}[c][c][1]{$E_{x}$}
\psfrag{k}[c][c][0.9]{$\boxed{\Delta\tau_1 = \Delta\tau_2\quad |T_{xx}^G| = -|T_{xx}^L|}$}
\psfrag{m}[c][c][0.9]{$\boxed{\Delta\tau_3 = 2\Delta\tau_{1,2}\quad |T_{xx}| \approx 0\text{dB}}$}
\includegraphics[width=0.8\columnwidth]{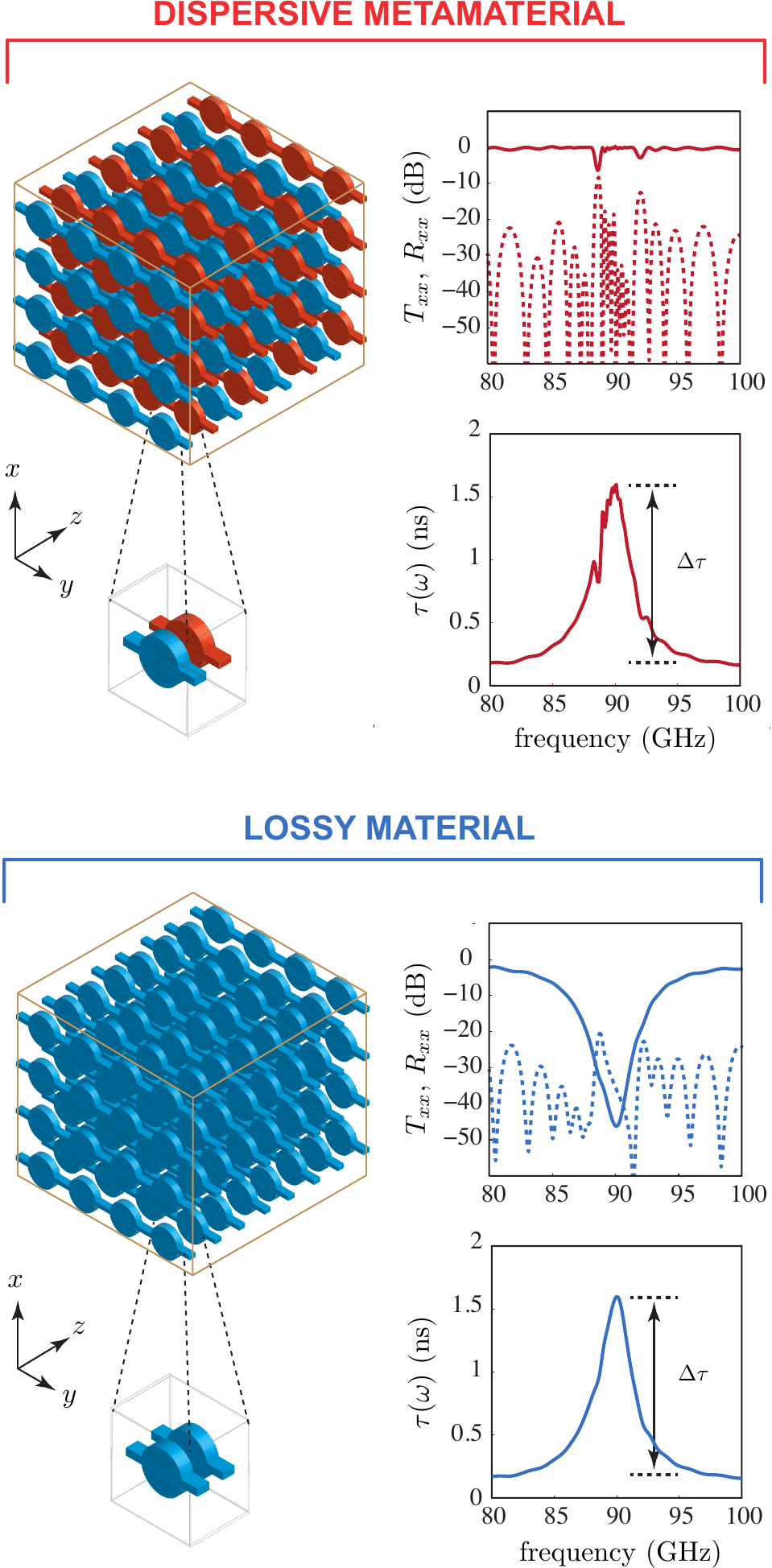}
\caption{Full-wave (RCWA) demonstration of a perfect dispersive Lorentz metamaterial formed by stacking $N=4$ of the perfect dispersive composite loss-gain metasurfaces in Fig.~\ref{Fig:Metasurface}(b), and comparison with a purely passive structure with the same number of layers.}
\label{Fig:Metamaterial}
\end{center}
\end{figure}

\section{Implementation of a Lorentz Perfect Metamaterial}

As mentioned in Sec.~\ref{eq:form_perf_disp_mtm}, a reflection-less metasurface may be realized using Huygens' scattering particles that satisfy the matching condition~\eqref{eq:matching_cond}. Circular-cylindrical dielectric resonators, which have been successfully incorporated in different types of metasurfaces~\cite{Shalav_Alldielectric,Kivshar_Alldielectric}, represent a convenient implementation of such particles. They also exhibit a Lorentz polarization response~\cite{Kerker_Scattering}, and are therefore appropriate to build Lorentz perfect dispersive metamaterials~(Sec.~\ref{sec:part_Lorentz_PDM}).

The structures to be considered in this section are based on 2D arrays of circular-cylindrical dielectric resonators mechanically held together by tiny dielectric bridges (playing only a minor electromagnetic role), as shown in the structures of Figs.~\ref{Fig:Metasurface} and~\ref{Fig:Metamaterial}. In these structures, loss and gain are included in the dielectric permittivity ($\epsilon_r = \epsilon' + i\epsilon''$) and the host medium is assumed to be air. The excitation is always a normally ($z$-directed) incident plane wave. All the results are full-wave simulation results, obtained by the Rigorous Coupled-Wave Analysis (RCWA) \cite{RCWA} technique, assuming infinite periodicity in the transverse ($xy$) plane.

Figure~\ref{Fig:Metasurface}(a) shows the structures and results for isolated loss and gain dielectric metasurfaces. The metasurfaces have been successfully designed to exhibit essentially the same dispersion responses as their ideal Lorentz counterparts in Figs.~\ref{Fig:Lorentzmedium}(a) and~(b), namely nearly identical group delays along with symmetric transmission magnitudes. Matching is not perfect, although it is still practically excellent ($R(\omega)<20$~dB). The little amount of reflection is due to imperfectly equalized electric and magnetic dispersions in the dielectric resonators. Quasi perfect matching may in principle be achieved using more sophisticated particle geometries, such as chiral ones~\cite{BBParticlesTratyakov}.

Figure~\ref{Fig:Metasurface}(b) shows the structure and results for the composite loss-gain metasurface formed by stacking together the electrically and magnetically matched loss and gain metasurfaces of Fig.~\ref{Fig:Metasurface}(b). The composite metasurface exhibits essentially the same dispersion response as its ideal Lorentz counterpart in Fig.~\ref{Fig:Lorentzmedium}(c), namely doubled group delay compared to the isolated loss and gain metasurfaces along with a flat 0~dB transmission magnitude, while preserving the excellent matching of the isolated metasurfaces.

Finally, a perfect dispersive Lorentz metamaterial formed by stacking perfect dispersive composite loss-gain metasurfaces is demonstrated in Fig.~\ref{Fig:Metasurface}(b). This metamaterial essentially reaches the goal set in Fig.~\ref{Fig:Definition}(c), namely providing high dispersion with nearly flat transmission and zero reflection. Assuming identical layers, as is the case in the figure, the group delay swing achieved in such a metamaterial is enhanced from $\Delta\tau_0$, where $\Delta\tau_0$ is the delay swing of a single metasurface (with loss or gain), to $N\Delta\tau_0$, where $N$ is the number of composites metasurfaces, while maintaining a nearly-flat transmission magnitude. For comparison, Fig.~\ref{Fig:Metamaterial} shows the response of a purely passive metamaterial with the same number of layers. While providing the same amount of dispersion, this medium suffers from a prohibitively large insertion loss ($\approx 50$~dB in the figure) at the group delay peak.

\section{Conclusions}

The concept of a perfect dispersive medium, which is a loss-gain medium characterized by a perfectly flat magnitude response along with an arbitrary phase (dispersion) response, has been proposed and realized in the form of an all-dielectric metamaterial. Such a medium has been described in terms of its electric and magnetic dispersions, which, when equalized, imply that the amplitude and phase of the transmission functions of the isolated gain and loss contributions invert and remain unchanged, respectively, under reversal of the sign of the imaginary part of the polarizability.  The proposed metamaterial has been demonstrated using a particular example of a stacked loss-gain metasurface structure formed using a 2-D array of subwavelength cylindrical dielectric resonators. Its flat magnitude response along with non-constant frequency dependent group delay has been confirmed using RCWA simulations.

A perfect dispersive metamaterial may be realized at radio frequencies using transistors and at optical frequencies using gain materials. However, this concept also pertains to simpler structures, such as for instance loss and gain optical microcring waveguide resonators coupled to a waveguide \cite{PT_microring,MicroRings}. The transfer function of a an isolated loss or gain waveguide ring resonator coupled to a straight waveguide is given by

\begin{equation}
T(\omega) = \left(\frac{\kappa\cos\gamma\ell - 1 + i\kappa\sin\gamma\ell}{\kappa\cos\gamma\ell - 1 - i\kappa\sin\gamma\ell}\right),
\end{equation}

\noindent with $\gamma = \beta - i\alpha$, where $\beta$ and $\alpha$ are the propagation constant and loss or gain coefficient, respectively, of the isolated waveguides, and $\kappa$ is the coupling coefficient between the ring and the waveguide and $\ell$ is the length of the coupling section. Substituting the complex propagation constant $\gamma$ in the above equation yields

\begin{equation}
T(\omega) = \left(\frac{\kappa e^{\alpha}\cos\beta\ell - 1 + i\kappa e^{\alpha}\sin\beta\ell}{\kappa e^{-\alpha}\cos\beta\ell - 1 - i\kappa e^{-\alpha}\sin\beta\ell}\right).
\end{equation}

\noindent It can be easily verified that $\angle T(\omega, \alpha)=\angle T(\omega, -\alpha)$ and $|T(\omega,\alpha)|= 1/ |T(\omega, -\alpha)$, which are the fundamental properties of a perfect dispersive system [Eq.~\eqref{Eq:Keyphasemag}]. Therefore, a cascade of coupled loss-gain ring resonators would form a 1D perfect dispersive structure exhibiting a flat transmission magnitude along with frequency dependent dispersion response.

The proposed perfect dispersive medium concept represents thus a practical solution to suppress the medium distortions that have been plaguing ultrafast signal processing systems for decades. It may therefore therefore benefit a myriad of novel ultrafast communication, sensing, imaging and instrumentation applications.

\bibliography{Gupta_Perfect_Dispersive_Medium_NatMat_2015}

\bibliographystyle{apsrev}

\section*{Supplementary Material}

\subsection*{I. Bode Gain-Phase Relation for a General Transfer Function with Complex Transmission Zeros}

Consider a transfer function $T(\omega)$ that is analytic and includes $N$ transmission zeros $\omega_n$ in the upper half of the complex $\omega$ plane. Due to the transmission zeros, corresponding to singular logarithm, the form~\eqref{Eq:Bode_rel_pair} of the Bode relation does not exist in this region. To resolve this issue, let us introduce the auxiliary complex function $T_\text{aux}(\omega)$ related to $T(\omega)$ according to

\begin{align}
T(\omega) &= T_\text{aux}(\omega) \prod_{n=1}^N(\omega - \omega_n) \notag\\
& =\underbrace{\prod_{n=1}^N(\omega_n^\ast - \omega)T_\text{aux}(\omega)}_{\tilde{T}(\omega)} \underbrace{\prod_{n=1}^N\frac{(\omega - \omega_n)}{(\omega_n^\ast - \omega)}}_{B(\omega)} \notag\\
& =\tilde{T}(\omega)B(\omega). \label{Eq:TmDecom}
\end{align}

\noindent In this relation, the function $\tilde{T}(\omega)$ is also analytic, as $T(\omega)$, but it has no zeros in the upper half plane since all the zeros of $T(\omega)$ have been transferred to the lower half plane. Therefore, its logarithm can be conventionally defined as

\begin{equation}
\ln \tilde{T}(\omega) = \ln |\tilde{T}(\omega)| + i\angle \tilde{T}(\omega).
\end{equation}

\noindent Moreover, $\ln \tilde{T}(\omega) \rightarrow 0$ as $|\omega|\rightarrow \infty$ since $\tilde{T}(\omega) \rightarrow 1$ as $T(\omega) \rightarrow 1$ in this limit because $|B(\omega)|=1$~$\forall \omega\in\mathcal{R}$. Therefore, $\ln\tilde{T}(\omega)$ is a complex function whose real and imaginary parts are related by the Kramers-Kronig relations~[Eq.~\eqref{Eq:KK}]. This leads to the magnitude-to-phase relation

\begin{equation}
\angle \tilde{T}(\omega) = -\frac{2\omega}{\pi}\mathcal{P} \int\displaylimits_{0}^{+\infty}
\frac{\ln |\tilde{T}(\omega')|}{\omega'^2 - \omega^2} d\omega'.\label{Eq:KKTAux}
\end{equation}

\noindent For $\omega\in\mathcal{R}$, $\ln |\tilde{T}(\omega)| = \ln |T(\omega)|$ and $\angle T(\omega) = \angle \tilde{T}(\omega) + \angle B(\omega)$. Substituting these relations into~\eqref{Eq:KKTAux} yields

\begin{equation}
\angle T(\omega) = -\frac{2\omega}{\pi}\mathcal{P} \int\displaylimits_0^{+\infty}\frac{\ln |T(\omega')|}{\omega'^2 - \omega^2} d\omega' + \angle B(\omega).
\end{equation}

%

\noindent This is the Bode relation between the transmission magnitude and phase of $T(\omega)$.

\subsection*{II. Effective Surface Polarizabilities of a Loss-Gain Metasurface Pair}

Consider a perfectly matched single loss-gain metasurface pair, as in Fig.~\ref{Fig:Lorentzmedium}(b). The corresponding Laplace transfer function is obtained by substituting $N=1$ and $j\omega=s$ in~\eqref{Eq:GDEa} and reads

\begin{align}
T_\text{m}(s) &= \overbrace{\left[\frac{1+s\eta_0\hat{\alpha}(s)/2}{1-s\eta_0\hat{\alpha}(s)/2}\right]}^{\text{Loss}}\overbrace{\left[\frac{1+s\eta_0\hat{\alpha}(-s)/2}{1-s\eta_0\hat{\alpha}(-s)/2}\right]}^{\text{Gain}}.
\end{align}

\noindent This transfer function can be alternatively expressed as the effective single-metasurface matched transmission transfer function (form of Eq.~\eqref{eq:Tomega})

\begin{subequations}
\begin{equation}
T_\text{m}(\omega) =\left[\frac{1+s\eta_0\hat{\alpha}_\text{eff}(s)/2}{1-s\eta_0\hat{\alpha}_\text{eff}(s)/2}\right],
\end{equation}
where
\begin{equation}
\hat{\alpha}_\text{eff}(s) = \frac{4\{\hat{\alpha}(s) + \hat{\alpha}(-s)\}}{4 + s^2\eta_0^2\hat{\alpha}(s)\hat{\alpha}(-s)}\label{Eq:Peff}
\end{equation}
\end{subequations}

\noindent is the effective polarizability.

Let us now introduce Lorentzian polarizabilities, as in Sec. III. Substituting~\eqref{Eq:Lorentz_dipolar_moment} into~\eqref{Eq:Peff}, one gets

\begin{equation}
\hat{\alpha}_\text{eff}^\text{Lorentz}(s) = 8A\omega_p^2\frac{(s^2 + \omega_0^2)}{4(s^2 + \omega_0^2)^2  + (A^2\eta_0^2\omega_p^2+4\gamma^2)s^2}.
\end{equation}

\noindent This expression can be further expressed in a simpler form using partial fractions as

\begin{equation}
\hat{\alpha}_\text{eff}^\text{Lorentz}(s) = \frac{c_1}{s^2 + \omega_1^2} +  \frac{c_2}{s^2 + \omega_2^2},\label{Eq:Alpha_Eff_s}
\end{equation}

\noindent with $c_1= 8A\omega_p^2(\omega_1^2-\omega_0^2)/(\omega_1^2-\omega_2^2)$,  $c_2= 8A\omega_p^2(\omega_0^2-\omega_2^2)/(\omega_1^2-\omega_2^2)$ and $\omega_{1,2}^2 = \kappa \pm 8\sqrt{\kappa^2 -\omega_0^4}$ with $\kappa = (A^2\eta_0^2\omega_p^2 + 8\omega_0^2+ 4\gamma^2)/8$. The effective surface polarizability~\eqref{Eq:Alpha_Eff_s} is related to its time-domain counterpart by the Laplace transform integral

\begin{equation}
\hat{\alpha}_\text{eff}(s) = \int_{-\infty}^{\infty}\tilde{\alpha}_\text{eff}(t)e^{-st}dt.\label{Eq:LT}
\end{equation}

\noindent Causality demands that $\tilde{\alpha}_\text{eff}(t) = 0$ for all times $t<0$, and therefore the above Laplace transform has a region-of-convergence (ROC) $\text{Re}\{s\} > 0$. Taking the inverse Laplace transform yields then the time-domain effective polarizability

\begin{equation}
\tilde{\alpha}_\text{eff}^\text{Lorentz}(t) = \left[\frac{c_1}{\omega_1}\sin(\omega_1 t) + \frac{c_2}{\omega_2}\sin(\omega_2 t)\right]U(t),
\end{equation}

\noindent where $U(t)$ is the unit step function, i.e. $U(t) = 1$ for $t\ge 0$ and $0$, otherwise \cite{Poon_KK}. The Fourier transform of $\tilde{\alpha}_\text{eff}^\text{Lorentz}(t)$ can now be computed as

\begin{equation}
\hat{\alpha}_\text{eff}^\text{Lorentz}(\omega) = \sum_{i=1,2}\frac{c_i}{\omega_i}\left[\frac{\omega_i}{\omega_i^2 - \omega^2}-\frac{i\pi}{2}\{\delta(\omega+\omega_i) - \delta(\omega-\omega_i)\}\right],	
\end{equation}

\noindent where it can be easily verified that the $\text{Re}\{\hat{\alpha}_\text{eff}^\text{Lorentz}(\omega)\}$ and  $\text{Im}\{\hat{\alpha}_\text{eff}^\text{Lorentz}(\omega)\}$ satisfy the Kramers-Kronig relations~\eqref{Eq:KK}.

\end{document}